\documentclass{article}

\PassOptionsToPackage{numbers, compress}{natbib}

\usepackage[]{neurips_2019}

\usepackage[utf8]{inputenc} 
\usepackage[T1]{fontenc}    
\usepackage{hyperref}       
\usepackage{url}            
\usepackage{booktabs}       
\usepackage{amsfonts}       
\usepackage{nicefrac}       
\usepackage{microtype}      
\usepackage{lipsum}
\usepackage{graphicx}
\usepackage{enumitem}

\title{Adaptive Reinforcement Learning Model for Simulation of Urban Mobility during Crises}

\author{%
  Chao Fan\thanks{contributed equally}, \space Xiangqi Jiang\samethanks, \space Ali Mostafavi\\
  Texas A\&M University\\
  College Station, TX 77843 \\
  \texttt{chfan@tamu.edu, amostafavi@civil.tamu.edu}
}

\begin{document}

\maketitle

\begin{abstract}
  The objective of this study is to propose and test an adaptive reinforcement learning model that can learn the patterns of human mobility in a normal context and simulate the mobility during perturbations caused by crises, such as flooding, wildfire, and hurricanes. Understanding and predicting human mobility patterns, such as destination and trajectory selection, can inform emerging congestion and road closures raised by disruptions in emergencies. Data related to human movement trajectories are scarce, especially in the context of emergencies, which places a limitation on applications of existing urban mobility models learned from empirical data. Models with the capability of learning the mobility patterns from data generated in normal situations and which can adapt to emergency situations are needed to inform emergency response and urban resilience assessments. To address this gap, this study creates and tests an adaptive reinforcement learning model that can predict the destinations of movements, estimate the trajectory for each origin and destination pair, and examine the impact of perturbations on humans’ decisions related to destinations and movement trajectories. The application of the proposed model is shown in the context of Houston and the flooding scenario caused by Hurricane Harvey in August 2017. The results show that the model can achieve more than 76\% precision and recall. The results also show that the model could predict traffic patterns and congestion resulting from to urban flooding. The outcomes of the analysis demonstrate the capabilities of the model for analyzing urban mobility during crises, which can inform the public and decision-makers about the response strategies and resilience planning to reduce the impacts of crises on urban mobility.\\
  
  \vspace{-.08in}
\end{abstract}

\section{Introduction}
The resilience of communities to crises such as flooding, wildfires, pandemics and other extreme events hinges on a deep understanding and effective simulation of the impacts of the crisis on human and physical environment \cite{Zhu2019}. Urban mobility plays a vital role in community resilience to crises by enabling populations to access critical facilities, such as healthcare, pharmacies, and grocery stores, for instance, \cite{Liu2019}; hence, the ability to simulate and examine the impacts of crises on urban mobility is essential to effectively improve the resilience of cities \cite{Sadri2020}. Studies, including Lei et al. \cite{Lei2020}, have examined the vulnerability and resilience of transportation systems during natural disasters and other crises. The majority of existing studies focus primarily on examining the vulnerability of physical road networks \cite{Fan2020e,Dong2020} and quantifying the effects of crises on traffic patterns \cite{Oh2020,Wang2020b}. A critical missing component is the capability to simulate crisis perturbation scenarios for predictive evaluation of the impacts on movement trajectories and traffic patterns at urban scale to inform emergency-response and resilience-planning decisions. To address this gap, the goal of this study is to create and test a deep learning model that can capture individuals’ movement patterns during normal situations and impose crisis-induced perturbations to simulate impacts on movement trajectories and traffic patterns. 

With the wide use of smartphones and apps with location services, a person’s digital footprints can be generated based on daily trips, which allow researchers to gather fine-grained data related to individuals’ movement trajectories during normal situations for model training and prediction \cite{Olmos2018}. Prior studies have attempted to address mobility simulation through statistical models and machine-learning techniques at different scales \cite{Yan2017}. One stream of the existing studies focuses on quantitative assessments of the dynamical and statistical properties of human travel \cite{Kitamura2000}. With more prevalence of location data related to mobile phone users, González et al. \cite{Gonzalez2008} demonstrated that human movement trajectories have a high degree of temporal and spatial regularity, as they follow simple reproducible patterns. This finding has significant implications for the predictability of individual mobility patterns in normal conditions. Furthermore, measuring the entropy of an individual’s trajectory, Song et al. \cite{Song2010} showed a 93\% potential predictability of human mobility under normal situations. The statistical metrics provide essential knowledge about the travel distance, radius, and frequently visited locations of humans \cite{Song2010a}. Despite these advances in the statistical characterization of aggregated human mobility patterns, limited models exist that could capture the temporal sequence of activities and spatial distribution of activities of each individual. The majority of the standard mobility models have limited capability for predicting individual mobility at a local scale, such as a neighborhood, in a specific timestamp. 

With advances in machine-learning techniques and enhanced computational power, it is now possible to learn and simulate finer-grained mobility patterns, such as the sequential patterns of individual location histories \cite{Tang2019}. In particular, deep neural networks (DNN) are prevalent in learning and simulating human dynamics and density from large data sets \cite{Li2017b}. For example, Wang et al. \cite{Wang2020a} designed a DNN architecture with an alternative-specific utility using behavioral knowledge to analyze human choices of their travel mode in normal circumstances. Hosseini et al. \cite{KhajehHosseini2019} introduced a deep learning-based methodology to predict traffic state using convolutional neural networks (CNN) by taking the individual vehicle-level data as inputs. The CNN model is further improved to capture citywide crowd activities with parameter efficiency and stability \cite{liang2020revisiting}. These deep learning techniques offer significant improvements in predicting individual mobility patterns with high resolution in terms of locations and timestamps \cite{Wang2020}. 

Human mobility, including origin-destination matrixes and the trajectories for each  origin and destination pair, however, is strongly influenced by the condition of road networks, specific gathering events, friendship effects \cite{Cho2011}, and residents’ travel habits and lifestyles \cite{Zong2019}. In particular, crises such as natural disasters cause perturbations in physical infrastructure and roads that influence human movement trajectories. For example, inundated road segments and buildings would necessitate evacuation of affected residents, thus reducing travel demand in certain areas \cite{Fan2020e}. Human trajectory data in the context of crises for model training and mobility simulation are scarce, as this data is rarely recorded in historical record \cite{Deville2014}, necessitating adaptive models for simulating human mobility under crisis scenarios (e.g., flooding, wildfires, and blizzards.) Information related to movement trajectories and traffic patterns in such extreme and complex situations is rarely recorded in historical data. The lack of data makes the simulation of  human mobility during crisis conditions challenging \cite{Lu2012}.

To address this information gap, this study tested an adaptive reinforcement learning model that can (1) learn individual mobility patterns from data collected from regular daily activities; and (2) simulate mobility under extreme events. The model incorporates both time and location factors as the input features and allows users (e.g., practitioners and decision makers) to adjust the environmental parameters for various application objectives in depending upon scenarios. To illustrate the application as well as the performance of the model in both prediction and application, we trained the model using data related to human mobility activities in Houston during March and April 2017, and used the model to simulate movement trajectory densities and traffic patterns during the flooding caused by the Hurricane Harvey in August 2017. The potential of the model for applications in emergency response and pandemic prediction is also discussed.

\section{Related work}
\subsection{Resilience of urban mobility to crises}
The role of urban mobility in a crisis such as access by emergency responders has been emphasized in the existing literature \cite{Wang2015a}. Assessing strategies to enhance the resilience of urban mobility is essential to mitigate the negative effects of crises and to improve the effectiveness of response efforts. To this end, existing studies \cite{Dong2019b,Wang2016} have put forth methodologies to evaluate the vulnerability of physical road networks and have proposed appropriate strategies for enhancement of urban mobility resilience. In another stream of work, Fan et al. developed a mathematical contagion model to predict the spread of floodwaters over road networks that informs about the disruptions of the urban mobility networks in flooding events \cite{Fan2020e}. Dong et al. proposed a machine-learning model that coupled road networks and water channels to examine the impacts of flooding on road networks through the lens of network dependencies \cite{Dong2020}. Despite the success of prior models, the majority of their efforts are focused on physical road networks, such as the accessibility to critical facilities. To expand the understanding of collective mobility patterns in crises, Wang et al. \cite{Wang2016} used empirical data to examine the impact of natural disasters on population travel distances and radius of gyrations. The result of their study indicates inherent resilience of urban mobility in crises. The majority of the existing studies (such as Lu et al., 2012 \cite{Lu2012}) have used empirical approaches or analytical methods (such as network science-based models) to examine mobility patterns when a population is hit by a crisis. A critical missing piece, however, is predictive data-driven models that can realistically simulate the spatial-temporal patterns of urban mobility during a crisis. Such predictive models and their resulting information should inform about the spatial distribution of the vehicles on road networks and emerging traffic jams caused by crisis perturbations, such as road inundations. This information is crucial for emergency response and mobility resilience enhancement; however, using empirical data from past crises, one could only evaluate one scenario of crisis impacts. The key to addressing this gap is a predictive data-driven model that can learn the patterns of movements by individual vehicles during normal situations and which is capable of imposing crisis-related perturbations to simulated changes in mobility patterns. Such models for urban mobility simulation in a crisis situation could offer a unique and effective approach that enables city planners, emergency managers, and decision makers to capture the spatial and temporal patterns of population movements during extreme events, and facilitate contingency and hazard mitigation planning.

\subsection{Mobility destination and trajectory prediction}
Simulating urban mobility is an important problem with a wide range of applications, such as traffic control, accident warning, pandemic prediction, and urban planning \cite{Chao2020}. Therefore, there is a need for general and robust methods to predict the destinations and trajectories for drivers, especially in populated urban areas \cite{Zhao2018a}. A number of prior studies have explored multiple methods to quantify, model and simulate human mobility, ranging from destination (next place) prediction \cite{Noulas2012} to route selection. This section discusses existing research work to show the achievements in addressing the challenge and to highlight the need for adaptive models to simulate urban mobility under crises situations. 

First, a number of existing studies have proposed and tested methods for next-place prediction \cite{Ma2013}. Next-place prediction, also called destination prediction in local urban scale, predicts the movement patterns of individuals and accordingly estimates the flow of population in both spatial and temporal manners. In next-place prediction, the majority of the methods deal with the pairs of origins and destinations based on individuals’ historical trip data. As mentioned earlier, statistical evidence shows the presence of regularity in human daily mobility patterns \cite{Song2010}. Hence, to detect and learn this regularity, one commonly adopted approach is to develop clustering algorithms such as k-means and DBSCAN (Density-based Spatial Clustering of Applications with Noise) algorithm based on historical travel data and to identify association rules between the origin and destination \cite{Chu2010}. Multiple studies \cite{Du2019} indicated that the clustering-based approaches have achieved very high performance in next-place prediction under normal circumstances. Location embedding is another method to capture location semantics with comprehensive numerical representations \cite{Cheng2013,WangandLi2017}. For example, Shimizu et al. proposed a place-embedding method that can learn fine-grained representation with spatial hierarchical information and which achieved higher accuracy in predicting the next place of human trips in urban areas \cite{shimizu2020learning}.

Second, in addition to single destination prediction, existing studies have also explored models for sequential patterns of individual travel to places \cite{Zhao2018}. In this task, Markov chain-based models, including Mobility Markov Chain \cite{gambs2012}, Mixed Markov Chain \cite{Asahara2011}, and Hidden Markov Model \cite{Mathew2012}, are commonly used to predict the location sequence of a given person/vehicle. Associated with location sequences but more related to route selection, other research is concerned with predicting the trajectory selected by an individual from an origin to a destination has garnered significant attention over the past few years \cite{Qiao2015}. The predicted trajectories allow researchers to not only estimate travel time \cite{Tang2018}, but also to analyze travel demand over urban road networks \cite{Dabiri2020}. Due to the complexity of trajectory prediction and required data scale, only a limited number of studies have examined this problem. The most commonly used approach is the shortest-path algorithm that constructs routable graphs from historical trajectory data and then computes the route based on travel frequencies \cite{wei2012}. This approach relies on traffic flow on specific road segments, but does not consider origin and destination pairs, that significantly influence route selection \cite{Yang2017}. More recent studies direct attention to the behavioral mobility patterns of individuals by inferring home and workplace, duration of activities. and other relevant temporal and spatial features for route prediction \cite{Alexander2015}. For example, Jiang et al. proposed a mechanistic modeling framework, called TimeGeo, that can generate mobility behaviors with a resolution of 10 minutes and hundreds of meters \cite{Jiang2016}. 

In summary, recent works have significantly advanced methods of simulating urban mobility and trajectory prediction under normal conditions based on the evidence of regularity in human movement and activity patterns. Still unsolved, however, is the challenge of adaptability of a model for simulating urban mobility when people are exposed to disruptive conditions, such as life-threatening situations during crises such as flooding, wildfires, and even pandemics. Modeling approaches with limited adaptability would not be useful in assessing movement trajectories and traffic under crisis situations. Advances in adaptive reinforcement learning models provide opportunities for simulating mobility under crises situations. In the following section, we will elaborate the proposed adaptive reinforcement learning model and then show its application in the context of flood impact analysis in Houston during Hurricane Harvey in 2017.

\section{Methodology}
\begin{figure*}
    \centering
    \includegraphics[scale=0.43]{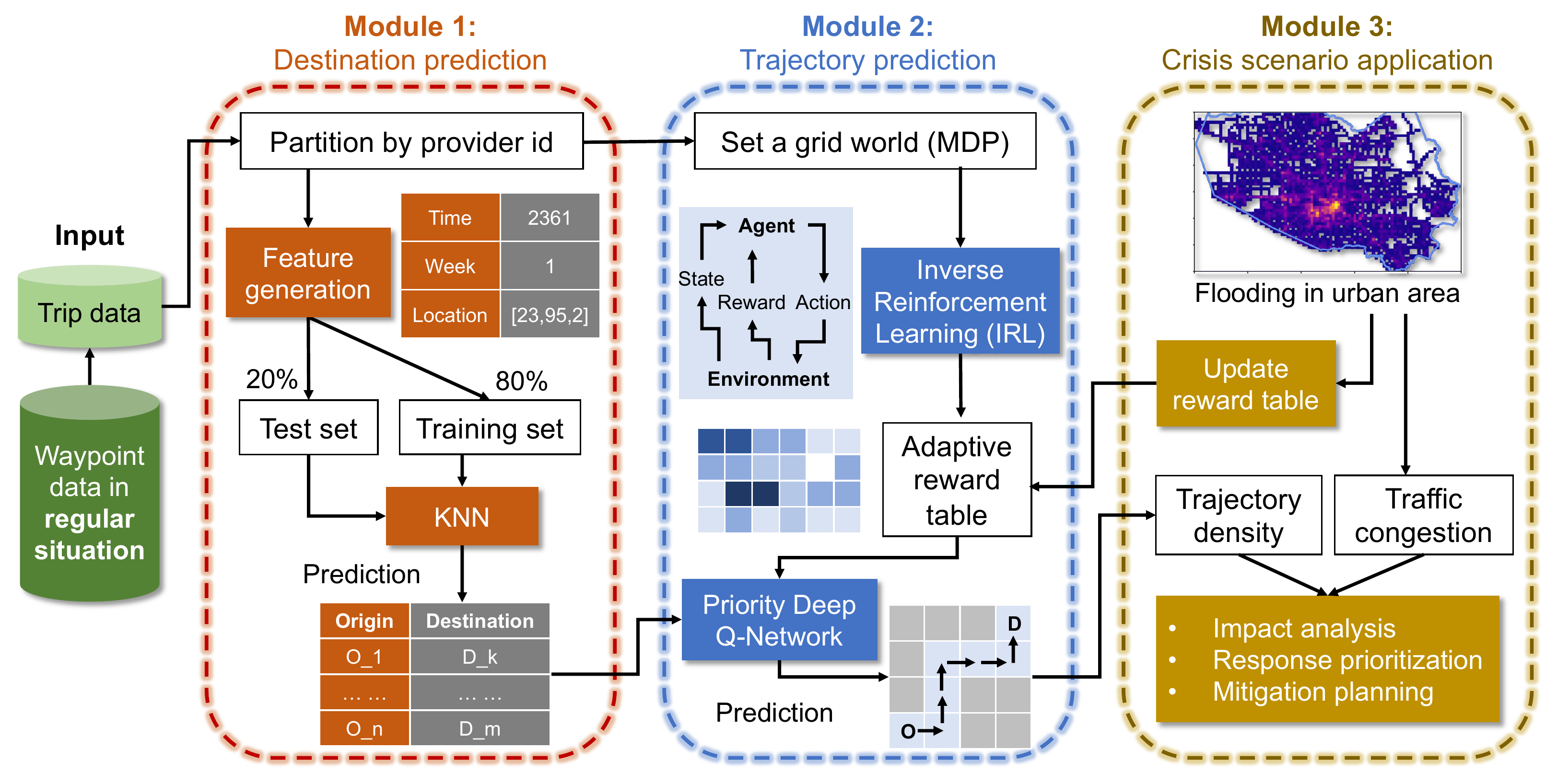}
    \caption{A schema of the proposed adaptive reinforcement learning model for simulating urban mobility.}
    \label{fig:1}
\end{figure*}

The proposed adaptive reinforcement learning model comprises three modules: destination prediction, trajectory prediction, and crisis scenario application (Figure \ref{fig:1}). The proposed model uses the human movement trajectory data as input, learns mobility patterns in regular situations, then simulates the trajectories and traffic conditions in crisis situations through adjusting the reward table. The mathematics and steps are formulated and elaborated in the following sub-sections.

\subsection{Preliminary definitions}
In this section, we first define the notations and terminologies that are used in this study.

\begin{itemize}[leftmargin=.3in]
  \item \textbf{Definition 1}. A \textbf{trajectory $T_i$}, also named as \textbf{waypoints}, \textbf{route} or \textbf{path}, is a spatial trace of a vehicle (or an individual) generated by their mobile apps in geographical space. $T_i$ contains a sequence of locations with specific latitudes and longitudes: $[(x_0^i,y_0^i,t_0^i),...,(x_{k^i}^i,y_{k^i}^i,t_{k^i}^i),...,(x_{n^i}^i,y_{n^i}^i,t_{n^i}^i)]$, where $x_{k^i}^i \in R$ and $y_{k^i}^i \in R$ are the latitude and longitude of a trace record $i$ at the $k^i_th$ location at time $t_{k^i}^i \in R$, and $k^i \in [0,1,...,n^i]$.
  \item \textbf{Definition 2}. A \textbf{trip} $t_i$, also called \textbf{O-D pair}, is a pair of origin and destination for a given trace record $i$. Trips can be extracted from trajectory data.
  \item \textbf{Definition 3}. A \textbf{provider} is a service provider which is allowed to anonymously collect locations from mobile devices. The providers are sorted into four categories: consumer vehicles, taxi/shuttle/town car services, field service/local delivery fleets, and for-hire/private trucking fleets. Each provider’s services fall exclusively into one category.
  \item \textbf{Definition 4}. A \textbf{grid world} is a grid based on a geographical map. The cells in the grid world are in the same shape and with equal area. Each cell represents a specific state (or location) showing is the location of a vehicle. This term is used to address the Markov decision process (MDP). 
  \item \textbf{Definition 5}. A \textbf{policy}, $\pi$, is a set of choices of actions for an agent (people or vehicle in this study) at each state.
\end{itemize}

Based on these definitions, the destination prediction problem can be described thus: given an origin with spatial coordinate $(x_0^i,y_0^i,t_0^i)$ and a specific scenario, predict the spatial coordinate of the destination $(x_{n^i}^i,y_{n^i}^i)$. The trajectory prediction problem can further be described as: given a pair of origin $(x_0^i,y_0^i,t_0^i)$ and destination $(x_{n^i}^i,y_{n^i}^i)$ with time information, predict a trajectory, $[(x_0^i,y_0^i,t_0^i  ),...,(x_{k^i}^i,y_{k^i}^i),...,(x_{n^i}^i,y_{n^i}^i)]$, that a vehicle (or an individual) will go through from the origin to the destination. The specific time in a location on the path to the destination is uncertain due to the dynamic traffic states and data collection frequency. Hence, in this study, the trajectory prediction task will predict only the route a vehicle (or an individual) will select from an origin to a destination, while the specific time for the location of the vehicle (or the individual) on the route will not be estimated.

\subsection{Feature generation}
The first module of the model is a tool that can predict the destination of a vehicle, given an origin at a specific time. As suggested by existing studies, human movements are spatial- and time-dependent, meaning that each pair of origin and destination should include both temporal and spatial features to ensure the model is both accurate and general. To achieve this, this section discusses the process of generating relevant features. 

For the temporal dimension, each trip sets the start time at the origin, showing the time of the day and day of the week. Empirical results show that urban mobility presents a significant weekly recurrent pattern \cite{Cho2011}. To incorporate this feature and to mitigate the effect of mobility variations across different days in a week, we converted the time of the day and day of the week into an integrated feature, called $t_n$. Specifically, we considered that any time within a week can be represented by a real number from 0 to 1 \cite{Garcia2017}. 0 represents the start of Monday (00:00 a.m., Monday), and 1 represents the end of Sunday (23:59 p.m., Sunday). Each day can be divided into 24 bins based on hours. Then, a specific time can be translated by:

\begin{equation}
   \centering
   t_n=\frac{hours+mins/60+24 \cdot weekday}{7 \cdot 24}
\end{equation}

where $hours$ represents the hours of the time in 24-hour system, $mins$ represents the minutes of the time, and $weekday$ represents the number of the days from Monday (for example, $weekday$=2 for Wednesday). In this way, we can specify the time feature of a trip in a continuous manner across different times in a week. 

Since people tend to have recurrent movement patterns during the weekday and special travels during the weekend (Saturday and Sunday), we also considered this difference in the model by using the category of weekday and weekend as a separate feature, $w_n$. The value of the trips in this feature can be represented using the following rule:

\begin{equation}
    \centering
    w_n=
    \left\{ \begin{array}{ll}
            0 & weekday \\
            1 & weekend
        \end{array} \right.
\end{equation}

The location of the origin and destination is usually recognized by the latitude and longitude coordinates. The resolution of the coordinates tends to be in meters. This raises a significant challenge for the predictive models to accurately learn the locations from the training data and predict location for a given origin. Also considering extremely high-resolution coordinates will also increase computational cost for the model. To simplify this process and maintain fairly high resolution, we round the coordinates into numbers with three decimals places, (This threshold is flexible and can be selected based on the requirement of the resolution.) Similar to the idea of grid, the coordinates are associated to a cell and represented by a centroid of the corresponding cell. Furthermore, since the earth is an ellipsoid which should be described as a three-dimensional space, we projected the latitude and longitude to a Euclidean 3D space using the following formula \cite{Garcia2017}:

\begin{equation}
    x_{st} = \cos(x_0^i) \cos(y_0^i), \\
    y_{st} = \cos(x_0^i) \sin(y_0^i), \\
    z_{st} = \sin(x_0^i)
\end{equation}

Through this approach, we can obtain three location features to represent a location of an origin $(x_{st},y_{st},z_{st})$. We also did the same transformation for the coordinates of the destination in the training data and obtain $(x_{ed},y_{ed},z_{ed})$.

\subsection{K-nearest neighbor regression for destination prediction and evaluation}
Once the five features for an origin are prepared, we can train a supervised learning model to learn the movement patterns by extracting the association between origins and destinations from the historical data in regular conditions. Among the commonly used learning methods \cite{Jia2017}, k-nearest neighbor (k-NN) regression demonstrates high performance in predicting the location feature values of the destinations. The k-NN regression method assigns weights based on the contributions of the neighbors for the predicted outcomes (location features in this study). The nearer the neighbors, the more contribution they can offer on the outcome values. 

Specifically, the k-NN model will take the list of trips $t_i=(O^i,D^i)$ as the training set, including all time and location features of origin and destination. $O^i=(t_n^i,w_n^i,x_{st}^i,y_{st}^i,z_{st}^i)$, and $D^i=(x_{ed}^i,y_{ed}^i,z_{ed}^i)$. Then, we defined the number ($k$) of nearest neighbors that would contribute to predicting the destination. The value of $k$ should be learned from the training data by minimizing the root mean square error (RMSE) which is defined below. The distance between every two neighbors (two origins: $O^i$ and $O^j$) is calculated using Euclidean norm ($L^2$ norm), $\parallel \cdot \parallel$, as shown below:

\begin{equation}
    \parallel O^i-O^j \parallel = \sqrt{(t_n^i-t_n^j)^2+(w_n^i-w_n^j)^2+(x_{st}^i-x_{st}^j)^2+(y_{st}^i-y_{st}^j)^2+(z_{st}^i-z_{st}^j)^2}
\end{equation}

The location feature values of the predicted destination are the average values of the location feature for the corresponding destinations of the k-nearest origins. This process generates three location feature values for the predicted destination. We then convert the location features back to the latitude and longitude with three decimal places of accuracy, analogous to the centroid of the destination cell. This coordinate is the predicted coordinates for the destination. By adjusting the value of $k$ and calculating corresponding RMSE based on all pairs of predicted coordinates and actual coordinates, we can find an optimized value of $k$ which can minimize the RMSE with high predictive performance. 

This process is analogous to addressing a regression problem. Hence, we use the RMSE \cite{Chai2014} to measure the error between the actual and the predicted values:

\begin{equation}
    RMSE=\sqrt{\frac{1}{N} \sum_{i}^{N} \frac{1}{2} [(x_{n^i}^i-{x'}_{n^i}^i)^2 + (y_{n^i}^i-{y'}_{n^i}^i )^2])}
\end{equation}

where ${x'}_{n^i}^i$ and ${y'}_{n^i}^i$ are the latitude and longitude of the predicted destination for a trace record $i$, and $N$ is the number of traces in the test data. Since we calculate the RMSE for each trace, the value of $N$, here, should be 1. The RMSE takes the Euclidean distances between the predicted outcomes and the observations which are then normalized across distances among all test data. 

\subsection{Markov decision process and reinforcement learning}
Once the destinations are predicted for given origins, the next step is to estimate the trajectories between the origins and destinations. An effective and commonly used approach is to structure the space of learned policies, which more generally called Markov decision process. The Markov decision process is a stochastic control process with a tuple of $(S,A,P,R)$, where $S$ is a set of states $s_i \in S$, here meaning the cell in a representing vehicle location; $A$ is a set of actions $a \in A$ which includes the four directions a vehicle can move from one cell to another in this context; $P({s'}_i | s_i,a)$ is the transition probability that action $a$ in state $s_i$ for agent $i$ leads to state ${s'}_i$ in the next timestamp. Following a transition matrix $T(s_i,a,{s'}_i)$; and $R(s_i,a,{s'}_i)$ is the reward function that specifies the reward a vehicle will immediately receive after transitioning from state $s_i$ to state ${s'}_i$ due to action $a$. 

To simplify the state space and also to enable the matrix computation, we create a grid world that represents all possible states of an agent in the environment. With a start state (i.e., the origin of a vehicle) and the terminate state (i.e., the destination), the objective of this MDP is to learn an optimal policy $\pi:S \rightarrow A$, that can maximize the total rewards for an agent to move in the grid world \cite{ziebart2008maximum}. To learn a policy, we first need to specify the parameters in the model, the probability distribution over a trajectory $P(\zeta | T,\kappa)$ ($\zeta$ is a trajectory; and $\kappa$ is the reward weights), transition matrix $T$, and the reward function $R$. The calculation starts with the transition matrix $T$. In this study, we assume that the agents have the same probability to go in each direction a among the four directions. Hence, the transition probability is set to be equal (the value is 1 if the action can be taken) among the four directions. Since the agent is usually not allowed to move out of the grid and the destination, we set the transition probabilities of some actions at the border and destination cell to be 0.

Then, the probability $P_T$ that an agent will be in the state ${s'}_i$ at next step given a state $s_i$, an action $a$, and a transition matrix $T$, can be computed as $P_T({s'}_i | a,s_i)$. The probability distribution over a trajectory $P(\zeta | T,\kappa)$ would be proportional to the product of all probabilities over the path:

\begin{equation}
    P(\zeta | T,\kappa) \propto \prod_{s_i,a,{s'}_i \in \zeta}  P_T({s'}_i | a,s_i)
\end{equation}

To learn the reward function from historical trajectory data, we maximize the likelihood of the historical trajectory data under the maximum entropy distribution:

\begin{equation}
    \kappa^* = \arg\max_\kappa \sum_{data} \log{P(\widetilde{\zeta} | T, \kappa)}
\end{equation}

where $\widetilde{\zeta}$ specifically refers to the historical trajectories. Solving this function by gradient-based optimization method, we can obtain the reward weights for the grid world. The reward value of a trajectory is simply the sum of the state rewards, which is calculated by the product of the path feature and the reward weights. 

\begin{equation}
    R(\mathbf{f}_{\widetilde{\zeta}})= \sum_{s_i \in \zeta} \kappa^\top \mathbf{f}_{s_i}
\end{equation}

where $\mathbf{f}_{\widetilde{\zeta}}$ is the path feature, which is the sum of the grid feature of each state, $\mathbf{f}_{s_i}$. The formula is shown as follows:

\begin{equation}
    \mathbf{f}_{\widetilde{\zeta}} = \sum_{s_i \in \widetilde{\zeta}} \mathbf{f}_{s_i}
\end{equation}

In implementing the model, the first challenge we encounter is the computational cost for training the agent on the grid world. That is because, in each step, when the agent wants to determine the action and next state, the model has to do matrix multiplications. Since we include the whole Houston metropolitan area to create the grid world, the matrix would be very large and, thus, calculating the reward would and informing the agent to make a decision would require an onerously long calculation time. This cost is usually not allowed for the model that wants to be applied to various contexts, especially in crisis settings. In fact, the cells distant from the origin and destination are not to make any impact on the trajectory selection of an agent. To reduce the computational cost, hence, in the implementation process, we draw only a small bounding box that can incorporate both origin and destination. Then, all cells in the bounding box are considered to be in the reward matrix to do reward calculations.

The second challenge would be the effect of the historical trajectories on the immediate reward values. As discussed earlier, the reward values in the reward table are learned from the historical trajectories that pass through the cell. For an origin and destination pair, we aim to find a trajectory to connect these two cells. The historical trajectories that connect other O-D pairs would, however, give weights to other cells in the reward table. Sometimes, the weights from the trajectories for other O-D pairs are quite high, inducing the agent to go other directions, moving around some high reward cells, and even leaving the actual destinations. To mitigate the effect of noisy historical trajectories, we include only the historical trajectories that pass through the given origin and destination to learn the reward table. This solution would effectively overcome the noise in the reward tables, reduce the training time, and finally improve the efficiency and performance of the model in learning reward tables. 

\subsection{Transformation of the reward function}
The reward table for each pair of origin and destination is the key to predicting the trajectory that an agent would select under a specific circumstance. Through examining the reward function learned from the Markov decision process, some common problems hinder an agent to find the optimal policy. 

First, due to the sparsity of the data, it is often the case that the immediate rewards among all cells are quite close to each other, meaning that an agent would be not able to effectively distinguish the benefits of two states to make an accurate choice of the next state. As a result, the model is usually unstable, generating different trajectories in different rounds of implementation. In addition, the reward values learned by the reinforcement learning model are always positive for all cells in the grid world. The agent tends to get stuck in some cells with relatively high rewards since the next state does not offer substantial reward to enable a moving action. To address these pitfalls, we conducted a normalization to project the values of the immediate rewards to a negative space with a range of $[-1,0]$. The normalized reward matrix is denoted as $R'$. The process can be formulated using the following equation:

\begin{equation}
    {r'}_{ij} = \frac{r_{ij}-r_{max}}{r_{max} - r_{min}}
\end{equation}

where $r_{ij}$ is the immediate reward in cell at $i^{th}$ row and $j^{th}$ column in the grid in according to reward function $R$; $r_{max}$ and $r_{min}$ are the maximum and minimum immediate reward respectively among all cells in the same grid. By doing so, the range of the reward values will be extended and distributed in a greater space. The negative values of the rewards can enable the agent to move out of some cells halfway on the trajectory and get a high reward only at the destination cell. 

Second, in the reward table, the reward of an action given to the agent does not always compel the agent to move towards the destination due to a high density of the road networks and concentration of historical trajectories. That means an agent tends to move around the cells with high reward values, sometimes even moving back to the origin. The agent, however, should intentionally move towards the destination. The closer the next state to the destination, the higher reward the agent can obtain. To this end, we created a new matrix, $L$, that can be added up to the reward table to compensate for the effect of proximity to the destination on trajectory selection. Empirically, the reward distribution accounts for the closeness to the destination centers at the destination cell. The reward has to be reduced if the agent does not get to the destination or if the agent moves further from it. The idea of the reward distribution is similar to the Gaussian distribution. Hence, we employed the Gaussian distribution to assign rewards to the cells in the grid world. The reward regarding the closeness to a destination is denoted as $l_{ij}$ for the cell at $i^{th}$ row and $j^{th}$ column. The value of $l_{ij}$ can be obtained using the following formula:

\begin{equation}
    l_{ij}=f(i)+f(j)-\delta
\end{equation}

\begin{equation}
    f(i)=  \frac{1}{\sigma \sqrt{2\pi}} e^{-\frac{1}{2}(\frac{i-\mu}{\sigma})^2}
\end{equation}

\begin{equation}
    f(j)=  \frac{1}{{\sigma'} \sqrt{2\pi}} e^{-\frac{1}{2}(\frac{j-{\mu'}}{\sigma'})^2}
\end{equation}

where $f(i)$ and $f(j)$ are the reward values calculated based on the row and column; $\mu$ and ${\mu'}$ are the row and columns of the destination, respectively; $\sigma$ and ${\sigma'}$ are calculated as the variance of rows and columns. $\delta$ is a parameter that controls the range of the reward values. Regarding the contribution of the closeness to the destination, the value of $\delta$ can be tuned to improve the prediction performance of the model.  

Third, the model usually does not converge at the destination cell because the reward at the destination is not significantly higher than other states. The agent would go back and forth around the destination cell. To address this problem, we simply assigned a large reward to the destination cell so that and the process can be converged when the agent arrives at and stops at the destination. The matrix can be created based on the formula shown as follows:

\begin{equation}
    d_{ij}=
    \left\{ \begin{array}{ll}
            500 & if \; ij \; is \; the \; destination \\
            0 & otherwise
        \end{array} \right.
\end{equation}

where $d_{ij}$ is the extra reward given to the destination cell, while other cell would receive no reward.

After all transformation matrices are prepared, finally, the reward table for a pair of origin and destination can be obtained by combining all of these matrixes:

\begin{equation}
    R_t = {R'} + L + D
\end{equation}

Using this transformed reward table, we can train agents for each origin and destination pair to effectively learn the optimal policy.

\subsection{Optimal trajectory prediction and evaluation}
Once the reward table for each pair of origin and destination is prepared, we define that the expected utility starting in $s_i$ and acting optimally is $V^* (s_i)$. The expected utility starting out having taken the action a from state $s_i$ and thereafter acting optimally is defined as $Q^*(s_i,a)$. Here $Q^*(s_i,a)$ is computed by:

\begin{equation}
    Q^*(s_i,a)= \sum_{{s'}_i} T(s_i,a, {s'}_i)[R_t (s_i,a,{s'}_i) + \gamma V^*({s'}_i)]
\end{equation}

where $\gamma$ is a hyperparameter that can be adjusted to balance the contributions of the reward function and the expected utility. Based on the historical data, we can maximize the value of a state by:

\begin{equation}
    V^*(s_i) = \max_a Q^*(s_i,a)
\end{equation}

Finally, we can learn the optimal policy: 

\begin{equation}
    V^*(s_i)= \max_a \sum_{{s'}_i} T(s_i,a,{s'}_i) [R_t (s_i,a,{s'}_i) + \gamma V^* ({s'}_i)]
\end{equation}

\begin{equation}
    \pi^*(s_i) = \arg\max_a \sum_{{s'}_i} T(s_i,a,{s'}_i)[R_t(s_i,a,{s'}_i) + \gamma V^* ({s'}_i)] 
\end{equation}

A deep Q-network (DQN) with prioritized experience replay is adopted to solve the behavioral policy and predict the optimal trajectory that the agent would select for a given O-D pair. Consider a historical trajectory where the model has learned and estimated the $Q^*$ value for an action. The empirical trajectories should be encouraged to be sampled in the prediction. To prioritize the empirical trajectories, we first measure the difference between the predicted $Q^*$ value and the experienced $Q^*$ value in the same state for the same action, which is represented by $\delta_i$:

\begin{equation}
    \delta_i = R_t(s_i,a,{s'}_i) + \gamma V_{\theta^-}^* ({s'}_i)-Q_\theta^*(s_i,a)
\end{equation}

where $\theta^-$ represents the target deep neural network, and $\theta$ represents the current deep neural network. The equation can further be reformulated as:

\begin{equation}
    \delta_i = R_t(s_i,a,{s'}_i) + \gamma Q_{\theta^-}^* ({s'}_i,\arg\max_a Q_\theta^*({s'}_i,a)) - Q_\theta^*(s_i,a)
\end{equation}

The left-hand side of the equation, $R_t(s_i,a,{s'}_i) + \gamma Q_{\theta^-}^*({s'}_i,\arg\max_a Q_\theta^*({s'}_i,a)))$, is the target value, and the right-hand side of the equation, $Q_\theta^*(s_i,a)$, is the predicted value. This equation provides a quantitative measure of how much the deep neural network can learn from the given experience sample $i$. This is notated as the Double-Q temporal difference (TD) error.

To find an appropriate $\theta$ and an optimal trajectory, the TD error is minimized by using expectation of the samples from the replay buffer $D$:

\begin{equation}
    \min_\theta E_{(s_i,a,R,{s'}_i)} \sim D[(R_t(s_i,a,{s'}_i) + \gamma Q_{\theta^-}^* ({s'}_i, \arg\max_a Q_\theta^*({s'}_i,a)) - Q_\theta^*(s_i,a))^2]
\end{equation}

By doing so, we can predict the optimal trajectory for each pair of origin and destination.

The performance of trajectory prediction is measured based on the overlaps between the actual and the predicted trajectory. Since we created the grid world for the Markov decision process and optimal policy learning, the trajectories are characterized by the cells. The intersection between the set of cells in the predicted trajectory and the set of cells in the actual trajectory indicates the precision and recall of the model for O-D pairs \cite{Buckland1994}. Hence, we adopted the following formulas for measuring the precision and recall for each O-D pair to assess the performance of the model:

\begin{equation}
    Precision = \frac{True \; positive}{True \; positive + False \; positive}
\end{equation}

\begin{equation}
    Recall= \frac{True \; positive}{True \; positive + False \; negative}
\end{equation}

where $True \; positive$ is the number passing cells that the model correctly predicts and that presents in the actual trajectory; $False \; positive$ is the number of passing cells that the model incorrectly predicts and that are not in the actual trajectory; $False \; negative$ is the number of cells that are incorrectly predicted to be out of the trajectory. The calculation of the precision and recall allows us to quantitatively assess the performance of the model for each O-D pair and to identify the specific correctly predicted cells. 

\subsection{Crisis scenario application with contextual factors}
The adaptability of the model is an important feature that enables the model to be widely used in various contexts, especially in crises situations, such as flooding, hurricanes, wildfires, and pandemics. The major change in the model due to the variation of the contexts is the destination and immediate reward when agents make their decisions on the actions. Therefore, in this section, we discuss the steps and strategies to update the destination selection and the reward table. The detailed steps are shown below:

\begin{itemize}[leftmargin=.3in]
    \item \textit{Step 1}: sampling the origins based on the historical data (or the density of population);
	\item \textit{Step 2}: predicting the destinations corresponding to each sampled origin using the well-trained first module of the model. In particular contexts, such as crises, some destinations might have been damaged or closed. Based on the observed situations, these destinations must be removed from results and only reachable destinations retained. 
	\item \textit{Step 3}: generating the situation matrix $F$ to represent the situation between the origin and destination and updating the transformed reward table with the situation matrix $F$. Take the flooding event as an example. The value of each cell $f_{ij}$ in the situation matrix $F$ can represent the extent of flooding in that cell, such as the number of flooded road segments or the area of the flooded cell. Then, multiplied by a parameter $1/\beta$ to normalize the values, we can add the matrix to the transformed matrix or subtract the transformed matrix with the situation matrix, following the formula shown below:
	\begin{equation}
	    R_t = R' + L + D - F/\beta
	\end{equation}
	where, the value of $\beta$ can be selected based on the contribution of the situation to the trajectory selection of the agents.
    \item \textit{Step 4}: predicting the trajectory of the given origin and destination using the updated reward table and the priority deep Q-network.
    \item \textit{Step 5}: augmenting the prediction results. It is important to note that the model can simulate only the trajectory for each origin and destination pair, instead of directly predicting the number of vehicles on the road. The model, however, has the capability to simulate traffic conditions on the road based on the simulated trajectories. In this step, the algorithm can first simulate the trajectory for an origin and destination pair 50 times. Then the algorithm selects the cells with high pass-through probability. We use the number of times, $n_{ij}$, that the cell is included in the predicted trajectory as the weight of the cell, $v_{ij}$, and other cells are considered as zero, as the equation shown below:
    \begin{equation}
        v_{ij}^{(k)}=
        \left\{ \begin{array}{ll}
            n_{ij}^{(k)} & if \; grids \; are \; selected \\
            0 & otherwise
        \end{array} \right.
    \end{equation}
    Each pair of origin and destination would have one vehicle matrix, $V^{(k)}$, that encompasses the entire grid world. Since a cell would likely to be on multiple trajectories for different pairs of origins and destinations, for all pairs of origins and destinations, we use $c_{ij}=\sum_{k=1}^{N} v_{ij}^{(k)}$  to represent the number of vehicles passing through this cell, where $N$ is the number of simulated pairs of origins and destinations. 
\end{itemize}

Through the implementation of all these steps, we can arrive at a final traffic matrix $C$ to represent the number of vehicles simulated in the cells on the grid world. This matrix would not only show the common trajectories that people would choose to reach their destinations, but also can inform the emergency response and resource allocation based on perturbations in mobility and traffic patterns in the context of crises. The capabilities of the model in simulation of urban mobility during crises is illustrated in the Section 4.4. 

\section{Results}
To demonstrate the performance and capabilities of the proposed model, we first introduced the data sets and then elaborated a specific use case of the model with an application in flooding impact analysis in the Houston metropolitan area in the context of Hurricane Harvey in 2017. 

\subsection{Data collection and preprocessing}
The data set, which is used to train, tune, and evaluate the adaptive reinforcement learning model, comes from INRIX, a private location intelligence company providing location-based data and analytics. The reason why we employ these data are twofold. First, the INRIX data provides very detailed coordinates along with the trajectory of the vehicles. That is, INRIX collected the coordinates of the vehicles every few seconds. Hence, the temporal and spatial resolution of the data is high. Second, the INRIX data are collected over the entire Houston metropolitan area over a time period of two months (March through April 2017) at all times of the day. This yielded a dataset of more than 26 million trip records collected in a continuous timeframe. Third, the INRIX data contains the trips for hundreds of providers (defined in Section 3.1) with four vehicle types. The vast volume of the trips and the diversity of vehicle types allow us to capture the various activity patterns and to enhance the robustness of the proposed approach in learning and simulating large-scale urban mobility. 

To prepare this data for training, we selected providers with large record sizes and discarded roughly 20\% of the trips that were too short (distance between origin and destination less than 5 miles) that might induce noise into the training process. Then, we randomly selected 80\% of the trips as the training set and 20\% of the remaining data as the testing set.  

\subsection{Destination prediction}
The first task is to predict the destinations, given the origins. We mainly train the model to automatically find an optimal value for the number of neighbors, $k$, which indicates the number of neighbors that should be included for estimating the coordinates of the destinations. Since we have a large number of providers with different types, we trained the model on the dataset from each provider separately. 

\begin{figure}
    \centering
    \includegraphics[scale=0.5]{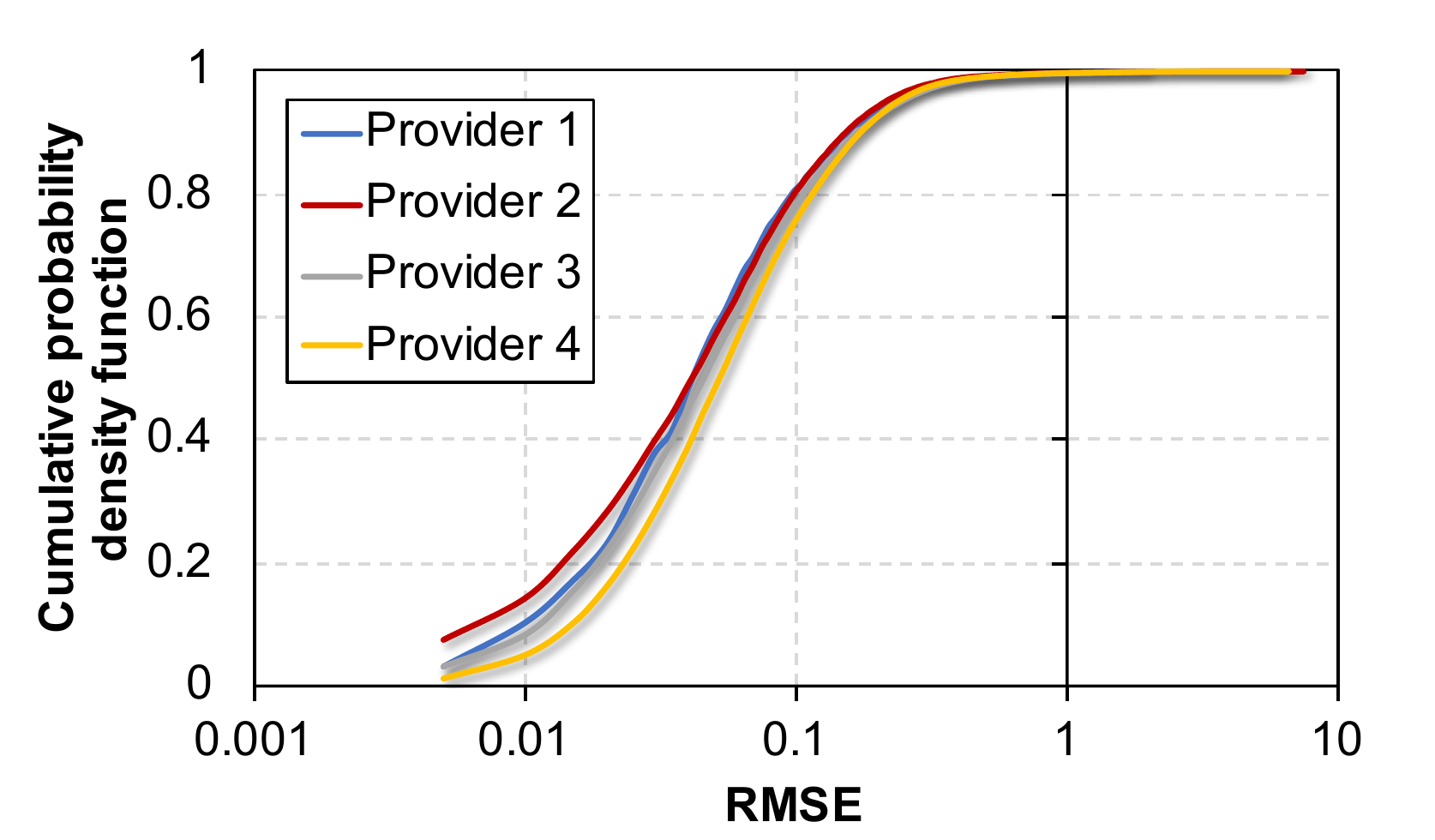}
    \caption{The performance of the model for destination prediction in four example providers. The average RMSEs for providers 1 through 4 are 0.0288, 0.0701, 0.0780, and 0.0824, respectively.}
    \label{fig:2}
\end{figure}

\begin{figure}
    \centering
    \includegraphics[scale=0.47]{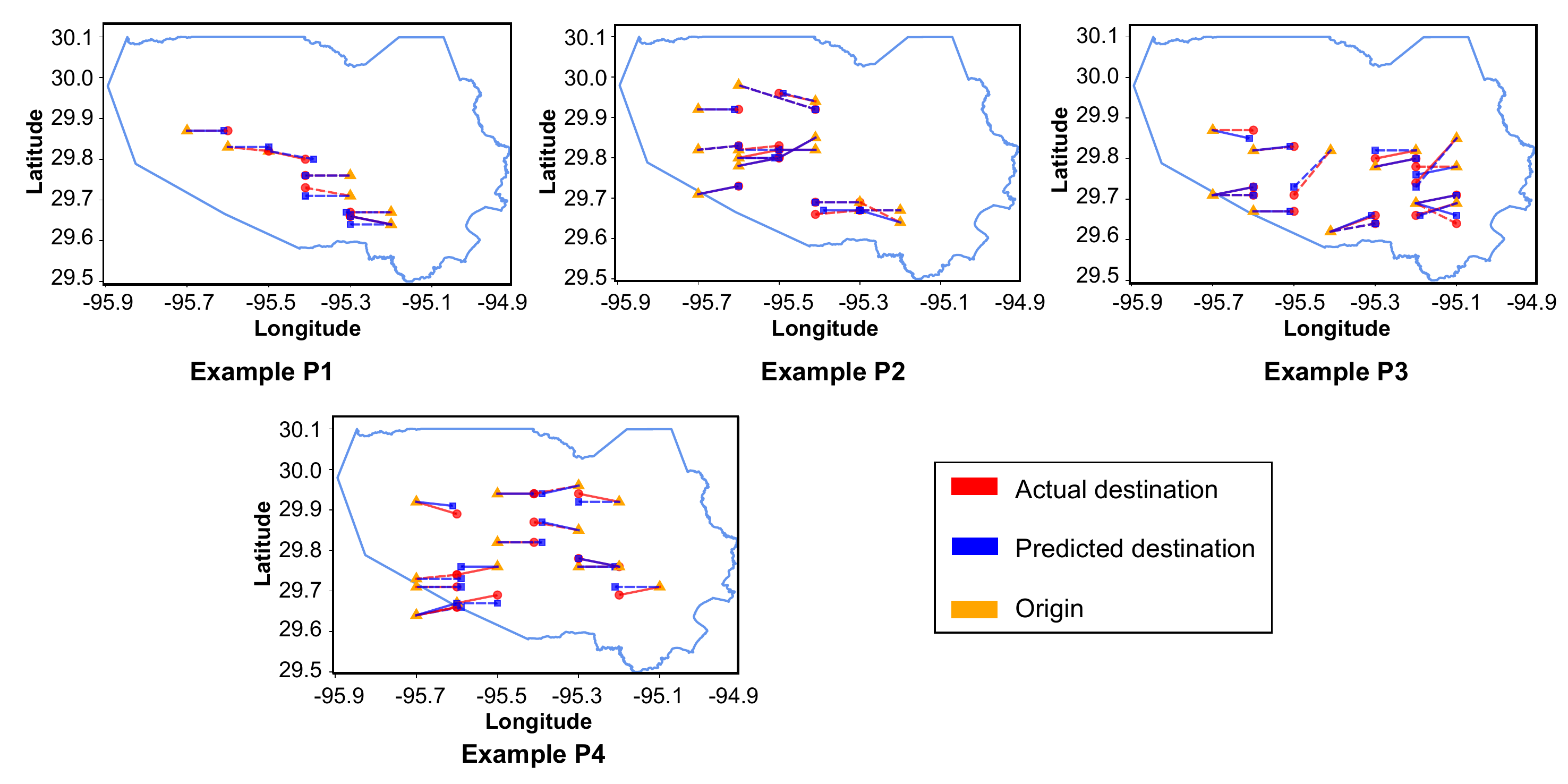}
    \caption{Example origin-destination pairs showing the performance of the model. (The blue polygon is the borderline of Harris County, Texas, USA.)}
    \label{fig:3}
\end{figure}

Figure \ref{fig:2} shows the prediction performance of the models in four example providers. As shown in the figure, the RMSE goes from a very small value, about 0.005, and then grows gradually to 0.1 when 80\% of the O-D pairs are predicted. Only 20\% of the O-D pairs have an RMSE greater than 0.1. The result indicates that the model can accurately predict the destinations for given origins based on the parameters learned from training data. Since we filtered out short-distance trips, the lengths of the remaining trips in the training and testing data are relatively long. By measuring the distances between the actual destinations and the predicted destinations, we find that a few predicted destinations are less than 0.3-mile from the actual destination, about 10\% of the predicted destinations are less than 0.6-mile from the actual destination, and more than 50\% of the predicted destinations are within the 3-mile distance from the actual destination. Considering the length of the trips themselves, the differences between the actual and predicted destinations are much shorter. Hence, the accuracy of the results is acceptable. Figure \ref{fig:3} further shows the spatial accuracy of some example O-D pairs. Figure \ref{fig:3} also shows the high performance of the model. Comparing the results among all providers, we also find that the model is quite stable with respect to the prediction performance and is not affected by the types of providers. This result demonstrates that the model is robust and generalizable to be applied to different data providers for destination prediction.

\begin{figure}
    \centering
    \includegraphics[scale=0.5]{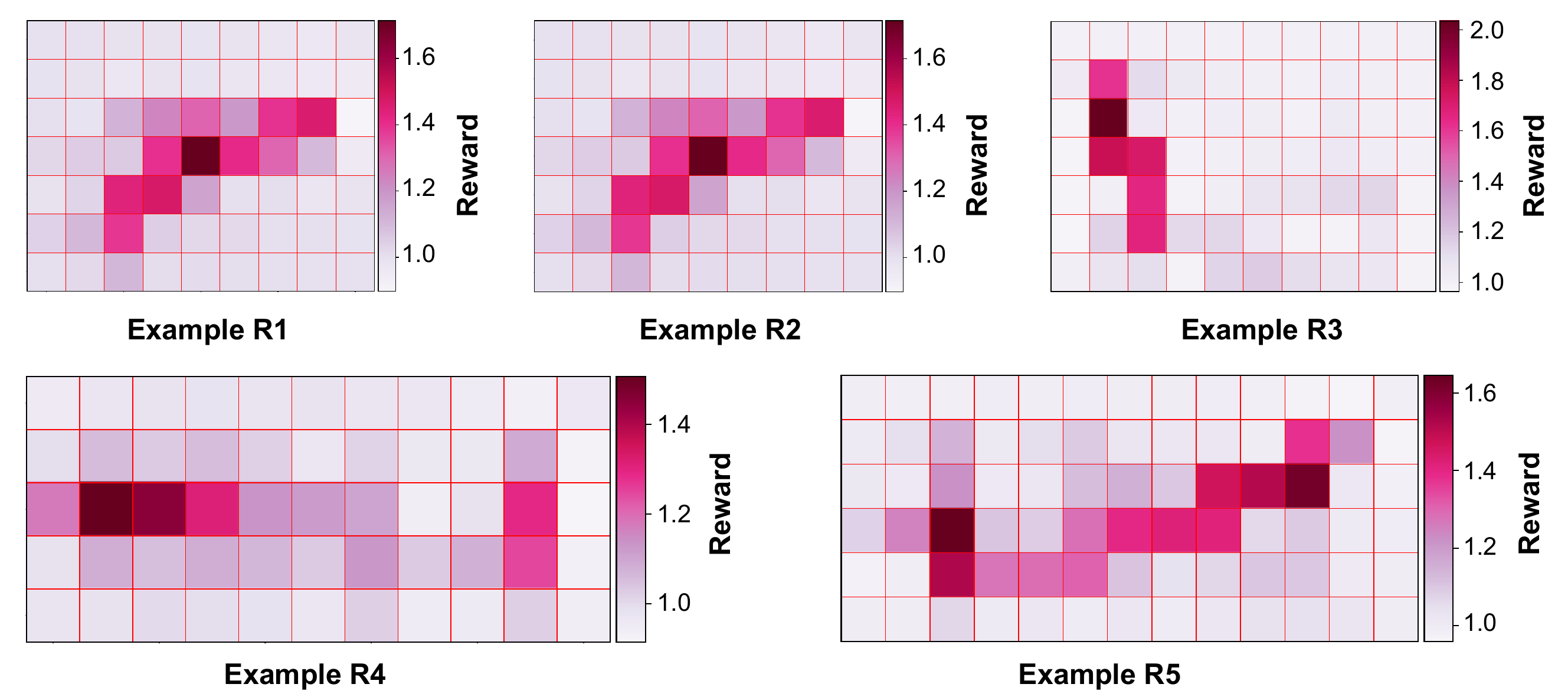}
    \caption{Example reward tables for origin and destination pairs, learned from training data using reinforcement learning.}
    \label{fig:4}
\end{figure}

\subsection{Trajectory prediction}
Once we acquire the destinations for given origins, we further use them to predict the optimal trajectories between the origins and destinations. The first step is to compute the reward table for each O-D pair. As discussed in the methodology section, we include only the cells within the bounding box. Figure \ref{fig:4} shows examples of reward tables for some of O-D pairs. The bounding box is a bit larger than the actual box tightly covering the origin and destination. In the majority of the example tables, such as R1, R2, and R5, we can find that the some significantly highlighted cells have high reward values, and the agent can effectively determine the optimal trajectories from the origin to the destination. Although historical data do not have a significant footprint for some O-D pairs, such as R3 and R4, the model can still impulse the agent to find an optimal trajectory. 

\begin{figure}
    \centering
    \includegraphics[scale=0.42]{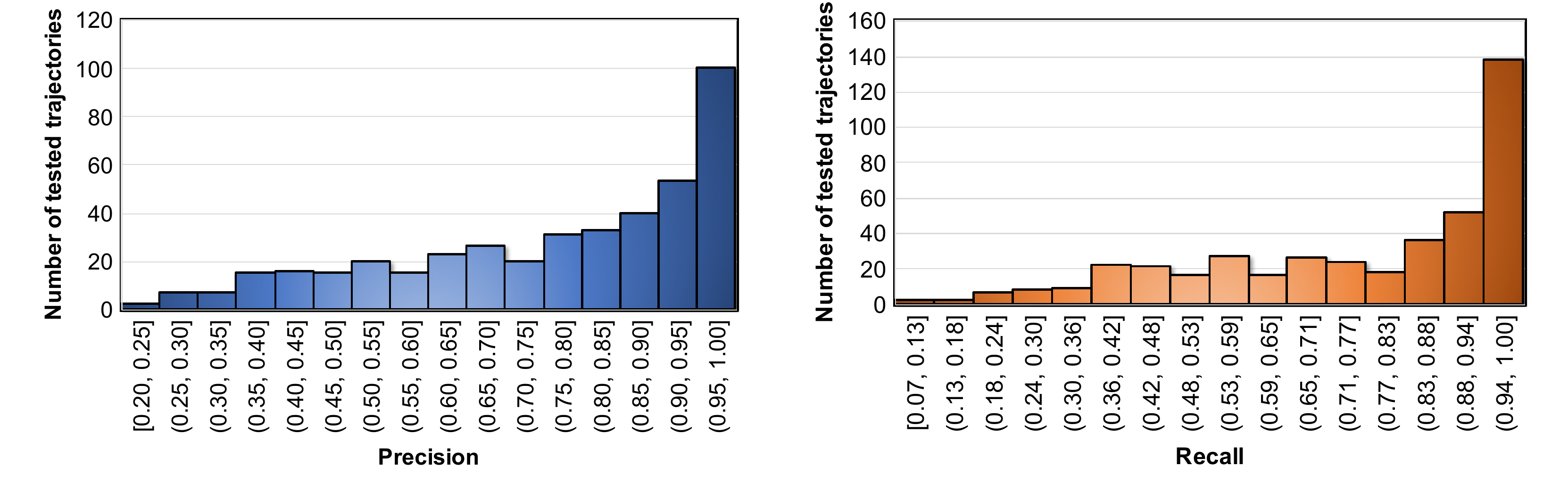}
    \caption{Performance of the trajectory prediction using module 2 of the proposed model (average precision: 0.765; average recall: 0.766).}
    \label{fig:5}
\end{figure}

\begin{figure}
    \centering
    \includegraphics[scale=0.41]{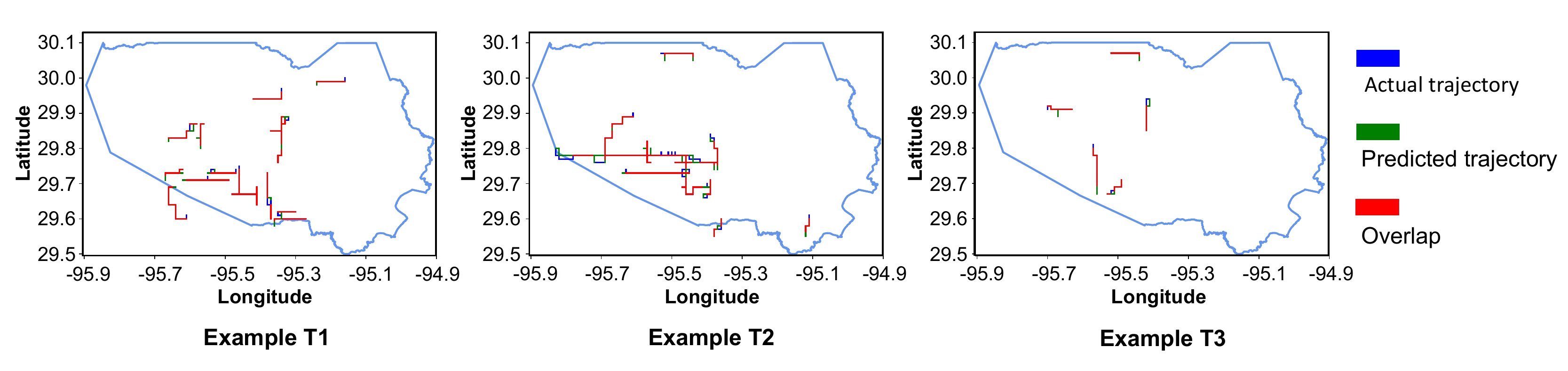}
    \caption{Example trajectories showing the performance of the model. (The blue polygon is the borderline of Harris County.)}
    \label{fig:6}
\end{figure}

A quantitative assessment of the model performance is shown in Figure \ref{fig:5}. We tested the model for multiple providers and O-D pairs. As the results suggest, the majority of the predicted trajectories are similar to the actual trajectories selected by the vehicles. Both average precision and recall are fairly high (although they are negatively affected by some extreme values). Figure \ref{fig:6} shows some examples of actual trajectories, predicted trajectories, and their overlaps. Despite the complexity of the historical trajectories transited by the vehicles, the model can still identify optimal trajectories for each pair of origin and destination. The performance of the model is also stable when it is trained and tested on different types of providers and datasets. These results and findings indicate that the proposed model is robust and could be used for trajectory-finding tasks.

\subsection{Simulation for flooding impact analysis}
To illustrate the application and capabilities of the proposed adaptive reinforcement model, we implemented the model for the situation during Hurricane Harvey in late August 2017 in the Houston metropolitan area. Hurricane Harvey, a Category 4 tropical storm, landed in Houston on August 26, 2017, and dissipated inland August 30, 2017. August 27 is the date with extreme and sustained rainfall that subsequently caused large-scale flooding over the urban areas, especially on urban road networks. As reported in news articles and government reports, more than 115,000 buildings were damaged, and 290 roads and highways were flooded \cite{Ibrahim2017,Sebastian2017}. To illustrate the performance of the model, we selected a specific short time interval, 10:00 a.m. through 12:00 p.m., August 29, 2017. We simulated vehicles that would move during this time interval and estimated their trajectory to understand the traffic conditions under the perturbation of urban flooding. Here we measured the traffic conditions in different locations by using the number of vehicles that move across a cell.

To be consistent, the data set used to validate the capability of the proposed model for scenario application was also obtained from INRIX. For each road segment, the dataset includes the average speed in 5-minute intervals and the speed in free-flow conditions (estimated as the speed limit for the roads). The flood situation on the road segments can also be reflected by the INRIX data. In the INRIX average speed dataset, the flooded road segments are identified by a designation of NULL for the traffic speed since no traffic records and no vehicles are driving through flooded road segments. To test the accuracy and validity of the data for identifying flooded roads, we also did a comparison of the NULL records between the datasets collected before and during the Hurricane Harvey, and also checked the records with flooded roads from government reports \cite{Dong2020}. We found that the NULL records presented only during the flooding period and were consistent with the flooded locations. 

\begin{figure}
    \centering
    \includegraphics[scale=0.5]{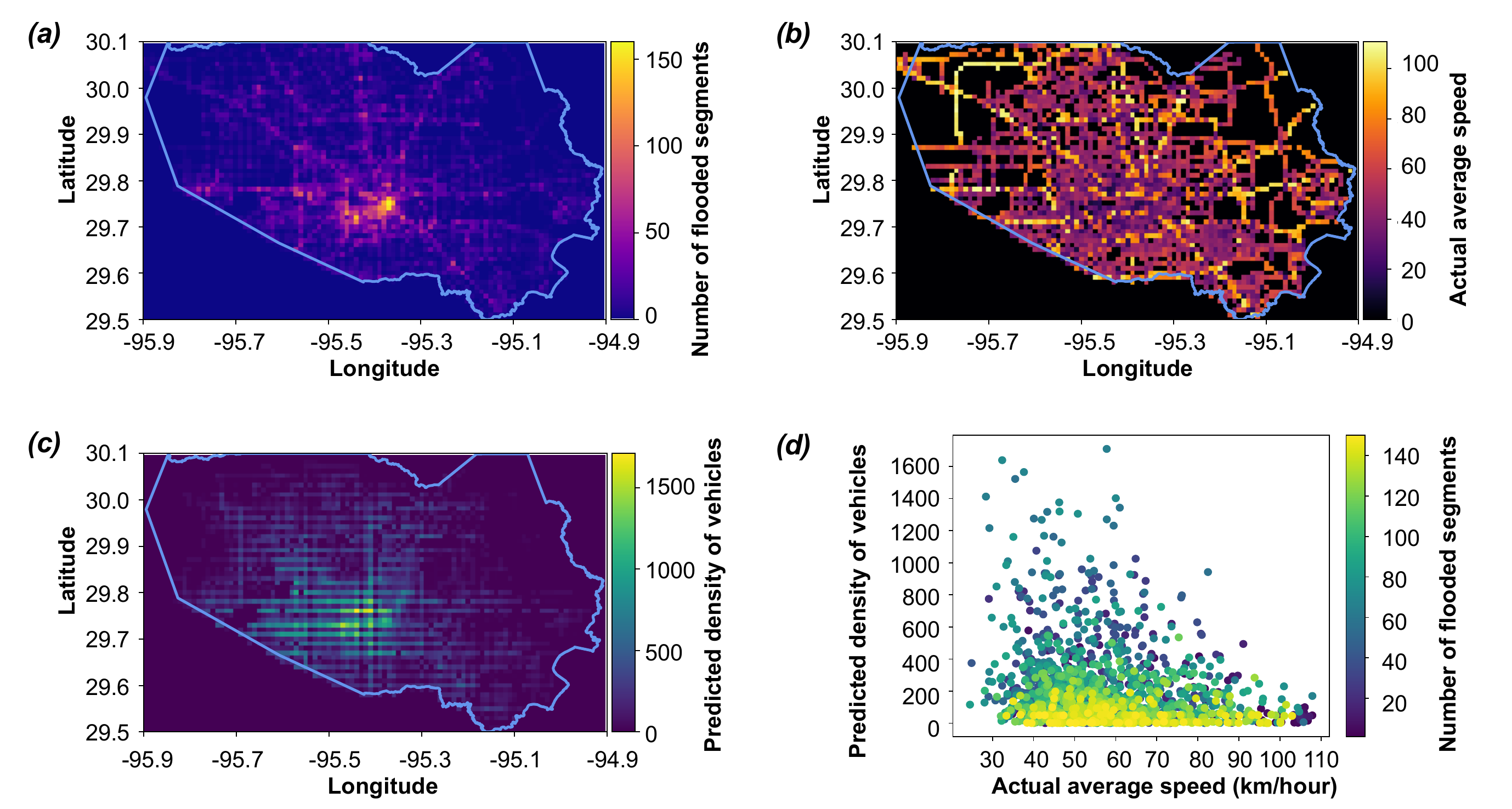}
    \caption{The simulation results. (a) Number of flooded segments in Houston, Harris County (the blue borderline); (b) Actual average speed for major road segments during the investigated time period; (c) Predicted density of vehicles during the investigated time period; and (d) Relationships among three variables for simulation validation.}
    \label{fig:7}
\end{figure}

Using this dataset, we identified the flooded road segments and quantified the extent of flooding in a cell. In the grid map for Houston, Harris County, each cell contains multiple road segments. To measure the extent of flooding in a specific cell, we considered the number of flooded road segments within a cell as a metric. As shown in Figure \ref{fig:7}(a), the flooded cells are concentrated on the central area of Houston and along the major roads such as highways. 

The traffic conditions in a cell can be captured by the actual traffic speed on the road segment. Since our model is mainly learned from the vehicles driving on major roads, here, we take into account only the road segments on which the speed limit is equal or greater than 50 km/hour so that all simulated results can be on the same road basis. This step also contributes to eliminating the effect of speed limits (types) of roads on the results. Hence, we computed the average of the actual speed on these selected road segments as the metric for traffic conditions for each cell. Figure \ref{fig:7}(b) shows areas where traffic was quite heavy due to flooding. 

To understand the impact of flooding on the population movement patterns, especially in the trajectory selection, we implemented the proposed adaptive reinforcement learning model by adopting the flood conditions to adjust the parameters in the model. The cells with more flooded road segments would have higher negative values added to the reward table. Accordingly, the agent would be less likely to transit the severely flooded cell to reach their destination. The simulated results are shown in Figure \ref{fig:7}(c) and (d). From these results, we found that the majority of the road segments hit hardest by the flood had fewer vehicles compared to the road segments where flooding was less severe (i.e., passable roads). This result indicates that, the model effectively simulated the trajectories selected by people who deliberately avoided flooded road segments. Hence, the road segments that were less flooded tend to have a great number of vehicles with low speed, or fewer vehicles with high speed. The results are quite realistic, which reveals the redistribution of vehicles and the utility of roads in urban road networks when flooding disrupted this area.

\section{Discussion and Concluding Remarks}
In this paper, we presented an adaptive reinforcement learning model that can simulate human mobility under crisis conditions based on the mobility patterns learned from historical data collected under normal circumstances. The proposed model can overcome the scarcity of mobility pattern data in crisis contexts, and also has the capability to simulate different crisis situations for impact analysis. The study showed the application of the proposed model using the digital trace data collected in Houston during regular situations and demonstrated the adaptability of the model in the case of urban flooding during Hurricane Harvey in 2017. 

The presented model has multiple contributions. First, the proposed simulation model and its adaptability to different crisis contexts can contribute to hazard mitigation and resilience improvement of urban areas. Predictive emergency warning and situational awareness are key components of hazard mitigation, which would benefit from effective simulation of crisis situations. Simulating urban mobility during crises is an essential step to capture the crisis situation and further inform early warning and mitigation planning. For example, using the proposed simulation model, city planners, emergency managers, and decision-makers can simulate scenarios of flooding (such as 100-year and 500-year flood scenarios) and road inundations to enable examination of the effects of movement trajectories and traffic congestion. This output information could inform about areas that would experience significant mobility disruptions and traffic congestion under varying flood scenarios. With sufficient examination of such contingencies, first responders and decision-makers can effectively assess the extent to which reducing the flood vulnerability of a major road or highway would decrease traffic congestion and improve movement trajectories. Also, the proposed model could be used to simulate mobility disruptions between critical origins (such as areas where vulnerable populations, such as the elderly and low-income households, are located) and critical destinations (such as healthcare facilities). In addition, the outcomes of the proposed model can help with near-real-time prediction of mobility disruptions and traffic congestions based on the real-time information, such as the inundated roads, to improve the accessibility of critical facilities and efficiency of responder dispatch. 

Second, the presented study also offers a new perspective for using deep learning techniques and digital trace data for urban mobility analysis. Although extremely big data related to human digital traces has been generated by daily mobility activity of individuals, existing deep learning techniques are limited to the prediction of regular mobility patterns. The proposed model enables consideration of contextual factors, such as road conditions, road network connectivity and population distribution, in the simulation process, which would significantly extend the capabilities of existing deep learning models gleaned from normal situations and applied to unfamiliar conditions, such as emergencies and traffic anomalies, to support decision-making and response strategies. For example, one can learn the mobility demand of the road networks from daily mobility data and predict the utility of each road segment by changing the layout of the networks. This can help design efficient transportation networks and traffic control systems to improve the performance of urban road networks. 

Finally, although the proposed model is general and can be adaptive to various phenomena and applications, future research could address some limitations. First, although this study achieves good performance for predicting the destinations and trajectories at a relatively fine-grained scale, the prediction of local movements, such as visits to nearby neighborhoods and points of interest, still needs to be improved. Local movements provide essential information about the lifestyle of residents. This study predicts the pure coordinates of destination and trajectories regardless of the specific locations visited and the roads traversed. Inclusion of this information in creating the movement profile of populations and predicting local movements would be of interest and importance in applications related to city and emergency planning. Second, the model has a great potential to be used as an application for rapid prediction. Implementing the proposed algorithm is still time-consuming, which would take a great amount of computational cost for rapid prediction. Hence, there is a need of improving the efficiency of the algorithm so that it can be used for more real-time prediction and across more cities and crises contexts.  

\section*{Acknowledgement}
This material is based in part upon work supported by the National Science Foundation under Grant Number CMMI-1846069 (CAREER), CMMI-1832662 (CRISP 2.0 Type 2), the National Academies’ Gulf Research Program Early-Career Research Fellowship, the Amazon Web Services (AWS) Machine Learning Award. The authors also would like to thank INRIX for providing the data for this research. Jan Gerston provided editorial services. Any opinions, findings, and conclusions or recommendations expressed in this material are those of the authors and do not necessarily reflect the views of the National Science Foundation and Amazon Web Services.

\section*{Data availability}
The data that support the findings of this study are available from INRIX and TTI (Texas A\&M Transportation Institute), but restrictions apply to the availability of these data, which were used under license for the current study, and so are not publicly available.

\section*{Competing interests}
The authors declare that they have no competing interests.

\section*{Author contribution}
C.F., X.J., and A.M. designed the study. X.J. and C.F. implemented the method and empirical case study. A.M. support with data acquisition. C.F. and A.M. wrote the main manuscript. All authors reviewed the manuscript.

\bibliographystyle{unsrt}  
\bibliography{Adaptive_RL_paper}

\end{document}